\documentclass[a4paper,11pt]{article}
\pdfoutput=1
\usepackage{jheppub} 
\usepackage[T1]{fontenc} 
\usepackage[english]{babel}
\usepackage[latin1]{inputenc}
\usepackage{amsmath,mathtools}
\usepackage{amsfonts}
\usepackage{amssymb}
\usepackage{wrapfig}
\usepackage{color}
\usepackage{float}
\usepackage{url}
\usepackage{mciteplus}
\usepackage{dsfont}
\usepackage{ifpdf}
\usepackage{epsfig}
\usepackage{xcolor}
\usepackage{hyperref}
\usepackage{cancel}
\usepackage{arydshln}
\usepackage{etoolbox}

\allowdisplaybreaks

\DeclareMathOperator{\Jt}{\widetilde{\mathnormal{J}}}

\DeclareMathOperator{\dt}{\widetilde{\mathnormal{d}}}

\DeclareMathOperator{\Rgod}{\mathnormal{R}_\mathnormal{g}\o\mathnormal{d}}
\DeclareMathOperator{\Rgodt}{\mathnormal{R}_\mathnormal{g}\o\widetilde{\mathnormal{d}}}
\renewcommand{\a}{\alpha}
\renewcommand{\b}{\beta}
\newcommand{\g}{\gamma}

\renewcommand{\k}{\kappa}
\renewcommand{\l}{\lambda}

\renewcommand{\L}{\mathcal{L}}
\newcommand{\M}{\mathcal{M}}
\newcommand{\N}{\mathcal{N}}
\renewcommand{\o}{\circ}
\newcommand{\Adg}{\text{Ad}_g}
\newcommand{\Adgi}{\text{Ad}_g^{-1}}
\newcommand{\Pm}{\mathnormal{P}_{-}}
\newcommand{\Str}{\text{Str}}
\newcommand{\tr}{\text{tr}}

\newcommand{\Rg}[1]{\mathnormal{R}_\mathnormal{g}\left(#1\right)}

\newcommand{\dds}{\mathnormal{d}^2\sigma\,}
\newcommand{\eRg}{\eta R_g}
\renewcommand{\d}{\delta}

\newcommand{\diag}{\text{diag}}

\newcommand{\algg}{\mathfrak{g}}
\newcommand{\alggb}{\mathfrak{g}_{\text{b}}}
\newcommand{\alggf}{\mathfrak{g}_{\text{f}}}

\newcommand{\so}{\mathfrak{so}}
\newcommand{\su}{\mathfrak{su}}
\renewcommand{\u}{\mathfrak{u}}
\newcommand{\uosp}{\mathfrak{uosp}}
\newcommand{\psu}{\mathfrak{psu}}
\newcommand{\algi}[1]{\mathfrak{g}^{(#1)}}
\newcommand{\Ai}[1]{A^{(#1)}}

\newcommand{\Z}[1]{\mathbb{Z}_{#1}}

\newcommand{\Em}[1]{E^{#1}}

\newcommand{\Cmn}[2]{C_{#1}^{\;\;#2}}
\newcommand{\Ctmn}[2]{\widetilde{C}_{#1}^{\;\;#2}}
\newcommand{\Lmn}[2]{\Lambda_{#1}^{\;\;#2}}
\newcommand{\dmn}[2]{\delta_{#1}^{\;\;#2}}
\newcommand{\Pb}[1]{P_2\left(#1\right)}
\newcommand{\jm}[1]{j^{#1}}
\newcommand{\jtm}[1]{\widetilde{j}^{#1}}
\newcommand{\Oi}{\mathcal{O}^{-1}}

\newcommand{\x}[1]{x^{#1}}

\newcommand{\xm}{x_{-}}
\newcommand{\xp}{x_{+}}

\newcommand{\dx}[1]{dx^{#1}}
\newcommand{\dxs}[1]{dx_{#1}^2}

\newcommand{\dxm}{dx_{-}}
\newcommand{\dxp}{dx_{+}}

\newcommand{\dxps}{dx_{+}^2}

\newcommand{\AdS}[1]{AdS_{#1}}
\newcommand{\Sph}[1]{S^{#1}}
\newcommand{\CP}[1]{\mathbb{CP}^{#1}}

\newcommand{\Sch}[1]{Sch_{#1}}

\newcommand{\vphi}{\varphi}

\newcommand{\ELm}[1]{\textbf{L}_{#1}}
\newcommand{\EMa}[1]{\textbf{M}_{#1}}
\newcommand{\EL}{\textbf{L}}
\newcommand{\EK}[1]{\textbf{K}_{#1}}
\newcommand{\EHi}[1]{\textbf{H}_{#1}}
\newcommand{\EH}{\textbf{H}}

\newcommand{\ET}[1]{\textbf{T}_{#1}}

\newcommand{\SOM}[1]{\textbf{M}_{#1}}
\newcommand{\Ep}[1]{\textbf{p}_{#1}}
\newcommand{\ED}{\textbf{D}}
\newcommand{\Epm}{\textbf{p}_{-}}
\newcommand{\Epp}{\textbf{p}_{+}}

\newcommand{\el}[1]{\lambda_{#1}}

\newcommand{\es}[1]{\sigma_{#1}}

\renewcommand{\th}[1]{\theta_{#1}}
\newcommand{\dth}[1]{d\theta_{#1}}
\newcommand{\vf}[1]{\varphi_{#1}}

\newcommand{\dvf}[1]{d\varphi_{#1}}
\newcommand{\dps}{d\psi}
\newcommand{\dji}{d\xi}

\newcommand{\Et}[1]{\eta_{#1}}
\newcommand{\fabc}[3]{f_{#1#2}^{\quad#3}}
\newcommand{\eabc}[3]{\epsilon_{#1#2}^{\quad#3}}

\newcommand{\gab}[1]{\gamma^{#1}}
\newcommand{\eab}[1]{\varepsilon^{#1}}

\newcommand{\Lap}{\nabla^2}

\title{\boldmath 
String Backgrounds of the Yang-Baxter Deformed $AdS_4\times\mathbb{CP}^3$ Superstring}

\author{Laura Rado,}
\author{Victor O. Rivelles}
\author{and Renato Sánchez}

\affiliation{Instituto de F\'{\i}sica, Universidade de S\~{a}o Paulo \\ Rua do Mat\~{a}o Travessa 1371, 05508-090 S\~{a}o Paulo, SP. Brazil}

\emailAdd{laura@if.usp.br}
\emailAdd{rivelles@fma.if.usp.br}
\emailAdd{renato@if.usp.br}

\abstract{
We build string backgrounds for Yang-Baxter deformations of the $\AdS{4}\times\CP{3}$ superstring generated by $r$-matrices satisfying the classical Yang-Baxter equation. We obtain the metric and the  NSNS two-form of the gravity dual corresponding to noncommutative and dipole deformations of ABJM theory, as well as a deformed background with Schr\"odinger symmetry. The first two backgrounds may also be found by TsT transformations while for the last background we get a new family of non-relativistic ABJM theories with Schr\"odinger symmetry.}

\begin{document}

\makeatletter
\patchcmd{\maketitle}{\@fpheader}{}{\hfill}{} 
\makeatother

\maketitle
\flushbottom

\section{Introduction}


Integrability is a very important feature found in several instances of the AdS/CFT correspondence. On the string theory side, formulated as a two-dimensional field theory, the notion of integrability is associated to the existence of a Lax connection which ensures the existence of an infinite number of conserved charges. In the case of $\AdS{5}\times\Sph{5}$ superstring, the theory is described as a $\sigma$-model on the supercoset $\frac{PSU(2,2|4)}{SO(1,4)\times SO(5)}$ \cite{Metsaev1998} and the $\Z{4}$-grading of the $\psu(2,2|4)$ superalgebra is a fundamental ingredient to obtain the Lax connection \cite{Bena2004}.


Recently, techniques to deform integrable theories while keeping their integrability have been developed. One of them is based on $r$-matrices that satisfy the Yang-Baxter equation. This kind of deformation was proposed by Klimcik as a way to obtain an integrable deformation of the Principal Chiral Model \cite{Klimcik2002,Klimcik2009}. He used a  
Drinfeld-Jimbo $r$-matrix \cite{Drinfeld1985,Jimbo1985} which satisfies the modified classical Yang-Baxter equation (mCYBE). This deformation was also applied to symmetric cosets $\sigma$-model \cite{Delduc2013} and furthermore to the $\AdS{5}\times\Sph{5}$ $\sigma$-model \cite{Delduc2014,Delduc2014a}. In this last case the supercoset construction was performed and the background was named $\eta$-deformed $\AdS{5}\times\Sph{5}$  \cite{Arutyunov2014,Arutyunov2015}. The important feature of this deformed background is that it does not satisfy type IIB supergravity field equations which led to a proposal for generalized supergravity equations \cite{Arutyunov2016,Wulff2016}. However, it was shown recently that the standard supergravity equations are satisfied by an $\eta$-deformed background if the Drinfeld-Jimbo $r$-matrix associated to this deformation is constructed in a specific form \cite{Hoare2019}.

It is also possible to consider $r$-matrices that are solutions of the classical Yang-Baxter equation (CYBE). The deformation of the symmetric coset $\sigma$-model was performed in \cite{Matsumoto2015} and for superstrings in $\AdS{5}\times\Sph{5}$ in \cite{Kawaguchi2014}. The interesting property of these deformations is that they lead to several known backgrounds of type IIB supergravity \cite{Matsumoto2014,Matsumoto2014a,Matsumoto2015a,Tongeren2015,Kyono2016}, namely Lunin-Maldacena-Frolov \cite{Lunin2005,Frolov2005}, Hashimoto-Itzhaki-Maldacena-Russo \cite{Hashimoto1999,Maldacena1999} 
\footnote{This background was also obtained as a  $\eta\rightarrow0$ limit of the $\eta$-deformed $\AdS{5}\times\Sph{5}$ background \cite{Arutyunov2014} after a rescaling \cite{Arutyunov2015,Hoare2016b},  showing that there are particular limits of $\eta$-deformed theories for which we obtain a standard supergravity solution.} and Schr\"odinger spacetimes \cite{Maldacena2008,Herzog2008,Adams2008}, which can also be obtained via TsT transformations \cite{Osten2017}. In these cases the $r$-matrices are all Abelian. These results were extended to the nonabelian case \cite{Orlando2016} and it was conjectured \cite{Hoare2016} that deformations using solutions of the CYBE are equivalent to nonabelian T-duality transformations \cite{Borsato2016,Borsato2017}. An interesting feature of these deformations is that they can generate partial deformations, i.e. either in $\AdS{5}$ or in $\Sph{5}$ for the $\AdS{5}\times\Sph{5}$ case. In terms of TsT transformations this can be explained by considering two-tori with directions either along the brane or transverse to it, or with one direction along the brane and the other transverse to it \cite{Imeroni2008}. In the first case the directions of the two-torus will be generated by momenta operators and this corresponds to a noncommutative two-torus in the field theory, i.e. a noncommutative field theory. By choosing one direction along the brane and one $U(1)$ direction, generated by a Cartan generator, transverse to the brane we get the so-called dipole deformation which gives nonlocal properties to the dual field theory without noncommutativity. If there are three $U(1)$ possible transverse directions, that is three Cartan generators, we can perform the deformation in three ways or more, generally a combination of them. This type of deformation breaks supersymmetry and Lorentz symmetry in the dual field theory. If the two-torus is in the transverse directions we get $U(1)\times U(1)$-type deformations called $\b$-deformations (or $\g$-deformation if the parameter is real). In general, having three $U(1)$ directions in the transverse space allow us to construct the transverse two-torus in three ways to get different deformations or a combinations of them. This type of deformations is marginal, i.e. it does not affect the conformal symmetry of the dual field theory but reduces the number of supersymmetries from $\N=4$ to $\N=1$. There is another way of getting a deformation from the TsT perspective: the null Melvin twist or TsssT transformation \cite{Bobev2009}. This is similar to the procedure for dipole deformations and can be extended to multiple parameters if the internal space has more than one $U(1)$ direction, as it is the case of $\Sph{5}$ and $\CP{3}$. In general, by deforming the $AdS$ subspace we expect, by the AdS/CFT correspondence, to obtain either nonconformal or noncommutative or  nonlocal theories as duals. On the other hand, by deforming the internal subspace we expect, in general, to reduce the supersymmetries on the field theory side. Thus, a combination of these deformations in the gravity side leads to interesting dual field theories. 

Another well-known case of the AdS/CFT correspondence is the duality between $\N=6$ superconformal Chern-Simons theory in three dimensions (or ABJM theory) and type IIA superstrings in $\AdS{4}\times\CP{3}$ \cite{Aharony2008}. The string theory is partially described by a $\sigma$-model on the supercoset $UOSp(2,2|6)/\left(SO(1,3)\times U(3)\right)$ \cite{Arutyunov2008,Stefanski2009} and since the superalgebra $\uosp(2,2|6)$ has a $\Z{4}$-grading it is possible to show its integrability \cite{Arutyunov2008}. Only recently Yang-Baxter deformations of the $\sigma$-model on this supercoset has been considered \cite{Negron2018}. A solution of the CYBE for an abelian $r$-matrix, in which only the $\CP{3}$ subspace was deformed, was found. This deformation leads to a three-parameter $\b$-deformation of the $\AdS{4}\times\CP{3}$ background that can be obtained also by using TsT transformation \cite{Imeroni2008}. Motivated by this work our aim is to study other possible integrable deformations of this background. 

The $r$-matrices satisfying the CYBE are constructed in terms of combinations of generators of the superalgebra $\uosp(2,2|6)$. We discuss $r$-matrices that lead to gravity duals of noncommutative ABJM theory as well as its three-parameter dipole deformation. These backgrounds were found initially by performing TsT transformations on the $\AdS{4}\times\CP{3}$ background \cite{Imeroni2008}. In addition to this, we also present $r$-matrices that lead to gravity duals of a nonrelativistic ABJM theory in a Schr\"odinger spacetime.

This paper is organized as follows. In \autoref{rmatrix} we present a short review on the construction of Yang-Baxter deformed $\sigma$-models and in \autoref{BCG} we derive the bosonic deformed background for a generic Yang-Baxter deformation.
In \autoref{cosetconstruction} we build the coset for $\AdS{4}\times\CP{3}$ paying attention to the relevant subalgebras that will be necessary in the following section. 
In \autoref{news} we discuss the NSNS sector of the new backgrounds obtained by deforming $\AdS{4}\times\CP{3}$ and in \autoref{conclusions} we discuss our results and future perspectives.

\section{Yang-Baxter Deformed \texorpdfstring{$\sigma$}{sigma}-models \label{rmatrix}} 

The action for the Yang-Baxter $\sigma$-model on a Lie superalgebra $\algg$ of a supergroup $G$ with $\Z{4}$ grading is given by \cite{Delduc2014} 
\begin{equation}\label{Sdef}
S=-\frac{\left(1+c\eta^2\right)^2}{2\left(1-c\eta^2\right)}\int \dds\Pm^{\a\b}\Str\left(A_\a,d\,J_\b\right),
\end{equation}
where the Maurer-Cartan one-form 
$A=g^{-1}dg\;\in\algg$, with  $g\in G,$ 
having $\Z{4}$-grading,  splits as
\begin{equation}
A=\Ai{0}\oplus\Ai{1}\oplus\Ai{2}\oplus\Ai{3},\quad \left[\Ai{k},\Ai{m}\right]\subseteq\Ai{k+m}\:\text{mod}\:\Z{4}.
\end{equation}
Also, $\Pm^{\a\b}=\frac{1}{2}\left(\g^{\a\b}-\kappa\varepsilon^{\a\b}\right)$, where $\g^{\a\b}$ is the worldsheet metric with $\det\g=1$, and $\k^2=1$ as required by kappa symmetry. The operator $d$ and its transpose\footnote{The operators $d$ and $\dt$ are such that
$\Str\left(M dN\right)=\Str\left(\dt MN\right)$ with $M,N\in\algg$.
} $\dt$ are
\begin{align}
\label{d1}
d&=P_1+\frac{2}{1-c\eta^2}P_2-P_3,\\
\label{d2}
\dt&=-P_1+\frac{2}{1-c\eta^2}P_2+P_3,
\end{align}
where $P_i$ $(i=1,2,3)$ are the projectors on each subalgebra of $\algg$ except $P_0$ on $\algi{0}$. This is required in order to have a $\algi{0}$-invariant action. The deformed current and its transpose are defined as
\begin{align}\label{J1}
J&=\frac{1}{1-\eta\Rgod}A,\\
\label{J2}
\Jt&=\frac{1}{1+\eta\Rgodt}A,
\end{align}
where the operator $R_g$ is 
\begin{equation}
\label{Rgdef}
\Rg{M}=\Adgi\o R\o\Adg\left(M\right)=g^{-1}R(gMg^{-1})g,\qquad g\in G.
\end{equation}
$R$ is the operator associated to the Yang-Baxter equation (YBE) which can be written, in its modified version, as
\begin{equation}\label{YBE}
\left[RM,RN\right]-R\left(\left[RM,N\right]+\left[M,RN\right]\right)=c\left[M,N\right],\left\{\begin{matrix}
c=0 & \text{CYBE}\\ 
c=\pm 1& \text{mCYBE}
\end{matrix}\right.
\end{equation}
where $M,N\in\algg$. In \eqref{Sdef} and \eqref{YBE} the parameter $c$ refers to either  the classical Yang-Baxter (CYBE) equation or to the modified classical Yang-Baxter equation (mCYBE).  

\section{Bosonic Backgrounds \label{BCG}}

In order to discuss the bosonic backgrounds we switch off the fermionic generators so that we now have 
$\alggb=\algi{0}\oplus\algi{2}.$
For the case of the CYBE we get from \eqref{d1} and \eqref{d2} that 
$d=\dt=2P_2$ 
where $P_2$ is the projector on $\algi{2}$. The deformed Lagrangian in \eqref{Sdef} reduces to
\begin{equation}
\label{YBdef}
\L=-\frac{1}{2}\left(\gab{\a\b}-\eab{\a\b}\right)\Str\left(A_{\a}\Pb{J_{\b}}\right),
\end{equation}
where we have chosen $\k=1$ for convenience.


Let us consider the case where $\algi{2}$ is a coset subalgebra $\algi{2}=\alggb/\algi{0}$. Thus, $\algi{0}$ has the generators of the local bosonic symmetries while $\algi{2}$ contains the generators which will allow us to construct the background. With this in mind, the projector $P_2$ can be written as
\begin{equation}
\label{P2def}
\Pb{X}=\sum_m\frac{\Str\left(\EK{m}X\right)}{S\tr\left(\EK{m}\EK{m}\right)}\EK{m}=X-\sum_i\frac{\Str\left(\EHi{i}X\right)}{\Str\left(\EHi{i}\EHi{i}\right)}\EHi,
\end{equation}
where $\EK{m}$ are the generators of $\algi{2}$, $\EHi{i}$ are those of $\algi{0}$ and $X\in\algg$.


The currents in \eqref{J1} and \eqref{J2} can be recast as 
\begin{align}
\label{ARgJ}
A&=\left(1-2\eRg\o P_2\right)(J),\\
\label{ARgJt}
A&=\left(1+2\eRg\o P_2\right)(\Jt),
\end{align}
and we can find $\Pb{J}$ by projecting \eqref{ARgJ} into $\algi{2}$,
\begin{align}
\label{P2A}
\Pb{J} =\Pb{A}+2\eta P_2\o R_g\o\Pb{J}.
\end{align}
We then find that $P_2$ acts as
\begin{equation}\label{PJK}
\Pb{A}=\Em{m}\EK{m}, \qquad \Pb{J}=\jm{m}\EK{m}, \qquad \Pb{\Jt}=\jtm{m}\EK{m}.
\end{equation}
In addition, we are going to need the following projection
\begin{equation}\label{Lmatrix}
\Pb{\Rg{\EK{m}}}=\Lmn{m}{n}\EK{n}.
\end{equation}
To do that we define from \eqref{J1}
\begin{equation}\label{Oidef}
\Oi=\frac{1}{1-2\eta R_g\o P_2},
\end{equation}
so that
\begin{equation}\label{POi}
\Pb{\Oi\left(\EK{m}\right)}=\Cmn{m}{n}\EK{n}.
\end{equation}
From \eqref{PJK}, \eqref{Lmatrix}, \eqref{Oidef} and \eqref{POi} we find 
\begin{equation}
\EK{m}=\Cmn{m}{n}\left(\dmn{m}{p}-2\eta\Lmn{m}{p}\right)\EK{p},
\end{equation}
or
\begin{equation}\label{Cmatrix}
\mathbf{C}=\left(\mathbf{I}-2\eta\mathbf{\Lambda}\right)^{-1}.
\end{equation}
Also, \eqref{P2A} becomes
\begin{equation}
\Em{n}\EK{n}=\jm{m}\left(\dmn{m}{n}-2\eta\Lmn{m}{n}\right)\EK{n},
\end{equation}
so that 
\begin{equation}\label{Emn}
\jm{m}=\Em{n}\Cmn{n}{m}, \qquad \jtm{m}=\Em{n}\Ctmn{n}{m},
\end{equation}
with 
\begin{equation}\label{Ctmatrix}
\widetilde{\mathbf{C}}=\left(\mathbf{I}+2\eta\mathbf{\Lambda}\right)^{-1}.
\end{equation}
Then, from \eqref{YBdef}, we can read off the metric and the B-field as 
\begin{equation}\label{Gmn}
ds^2=\Str\left(A\,\Pb{J}\right)=\jm{m}\Str\left(A\EK{m}\right)=\Em{m}\Cmn{m}{n}\Str\left(A\EK{n}\right),
\end{equation} 
\begin{equation}\label{Bmn}
B=\Str\left(A\wedge\Pb{J}\right)=-\jm{m}\wedge\Str\left(A\EK{m}\right)=\Em{m}\Cmn{m}{n}\wedge\Str\left(A\EK{n}\right).
\end{equation}

\section{Coset Construction for \texorpdfstring{$\AdS{4}\times\CP{3}$}{AdS4xCP3} \label{cosetconstruction}}

Arutyunov and Frolov \cite{Arutyunov2008} and Stefanski  \cite{Stefanski2009} 
proposed a supercoset $\sigma$-model formulation for type IIA superstrings in $\AdS{4}\times\CP{3}$ along the same lines as what it was done for type IIB superstrings in $\AdS{5}\times\Sph{5}$.
In this formulation type IIA superstring theory in $\AdS{4}\times\CP{3}$ is described by a $\sigma$-model on the supercoset $UOSp(2,2|6)/\left(SO(1,3)\times U(3)\right)$.
However, in this supercoset the spinors have 24 components instead of the usual 32 components so that 
it does not describe the full superstring. This means that the missing components have been gauged away so that this coset has a partially fixed $\kappa$-symmetry \cite{Arutyunov2008}.

The isometry group of $\AdS{4}\times\CP{3}$ is the coset
\begin{equation}
\AdS{4}\times\CP{3}\equiv\frac{SO(2,3)}{SO(1,3)}\times\frac{SU(4)}{U(3)}.
\end{equation}
This bosonic group is part of the supercoset $UOSp(2,2|6)/\left(SO(1,3)\times U(3)\right)$. The supergroup $G=UOSp(2,2|6)$ has superalgebra $\algg=\uosp(2,2|6)$ on which the $\sigma$-model can be constructed. The generators of $\algg$ can be written as supermatrices formed by blocks that correspond to bosonic and fermionic generators.
The bosonic part of $\algg$ can be written as
\begin{equation}
\alggb:=\so(2,3)\oplus\su(4)=\overbrace{\left(\so(1,3)\oplus\u(3)\right)}^{\algi{0}}\oplus\overbrace{\left(\frac{\so(2,3)\oplus\su{4}}{\so(1,3)\oplus\u(3)}\right)}^{\algi{2}}.
\end{equation}
Then a  supermatrix on $\algg$ then has the following form 
\begin{equation}
M_{(4|4)}=\left(\begin{array}{c|c}
\so(2,3) & \overline{Q}\\  
\hline
Q & \su(4)
\end{array}\right),
\end{equation}
where $Q,\bar{Q}$ are the fermionic blocks corresponding to the fermionic sector of $\algg$, $\alggf=\algi{1}\oplus\algi{3}$. As discussed in \cite{Negron2018}, in order to get the Fubini-Study metric for $\CP{3}$, we have to  extend the coset of $\CP{3}$ with an $\su(2)$ algebra. Then the coset for $\AdS{4}\times\CP{3}$ is now written as
\begin{equation}
\label{algB}
\alggb:=\so(2,3)\oplus\su(2)\oplus\su(4)=\overbrace{\left(\so(1,3)\oplus\su(2)\oplus\u(3)\right)}^{\algi{0}}\oplus\overbrace{\left(\frac{\so(2,3)\oplus\su(2)\oplus\su{4}}{\so(1,3)\oplus\su(2)\oplus\u(3)}\right)}^{\algi{2}}.
\end{equation}
The supermatrix for  $\algg$ will then have the following structure
\begin{equation}
M_{(6|4)\times(6|4)}=\left(\begin{array}{c:c|c}
\so(2,3) & 0 & \overline{Q}\\ 
\hdashline
0 & \su(2)& 0\\ 
\hline
Q& 0 & \su(4)
\end{array}\right),
\end{equation}
where the dashed line splits the algebras corresponding to the subspaces $\AdS{4}$ and $\CP{3}$, and the solid line splits the  $M_{6\times 6}$ and $M_{4\times 4}$ bosonic blocks. Notice that this extension does not mix with the original fermionic blocks. 

The basis of $\so(2,3)\oplus\su(2)\oplus\su(4)$ that we will consider is composed of $\so(2,3)$ generators denoted by $\SOM{ij}$, $i,j=0,1,2,3,4$, $\su(2)$ generators denoted by $\EMa{a}$, $a=1,2,3$, and $\su(4)$ generators denoted by $\ELm{m}$, $m=1,\dots,15$,
\begin{equation}
\label{supergens}
\SOM{ij}=\left(\begin{array}{c:c|c}
m_{ij} &  & \\ 
\hdashline
 & 0 & \\ 
\hline
 &  & 0
\end{array}\right),
\EMa{a}=-\frac{i}{2}\left(\begin{array}{c:c|c}
0 &  & \\ 
\hdashline
 & \es{a} & \\ 
\hline
 &  & 0
\end{array}\right),
\ELm{m}=-\frac{i}{2}\left(\begin{array}{c:c|c}
 0&  & \\ 
\hdashline
 & 0 & \\ 
\hline
 &  & \el{m}
\end{array}\right),
\end{equation}
where $m_{ij}$ are the ten $4\times 4$ antisymmetric matrices representing the generators of isometries of $\AdS{4}$ and $\es{a}$ and $\el{m}$ are, respectively, the conventional $2\times 2$ Pauli and $4\times 4$ Gell-mann matrices of $\su(2)$ and $\su(4)$ (see \autoref{ApSU4} and \autoref{ApSO23}). The commutation relations and supertraces are
\begin{equation}
\begin{gathered}
\left[\SOM{ij},\SOM{k\ell}\right]=\Et{i\ell}\SOM{jk}+\Et{jk}\SOM{i\ell}-\Et{ik}\SOM{j\ell}-\Et{j\ell}\SOM{ik},\\
\Str\left(\SOM{ij}\SOM{k\ell}\right)=-\Et{ik}\Et{j\ell},
\end{gathered}
\end{equation}
with $\Et{ij}=\diag\left(-,+,+,+,-,+,+,+,+,+\right)$, and
\begin{align}
\left[\ELm{m},\ELm{n}\right]&=\fabc{m}{n}{p}\ELm{p}, \qquad \Str\left(\ELm{m}\ELm{n}\right)=\frac{1}{2}\d_{mn},\\
\left[\EMa{a},\EMa{b}\right]&=\eabc{a}{b}{c}\EMa{c}, \qquad \Str\left(\EMa{a}\EMa{b}\right)=-\frac{1}{2}\d_{ab}.
\end{align}

The global symmetry algebra of $\AdS{4}$ space can be written as 
\begin{equation}
\so(2,3)=\so(1,3)\oplus\frac{\so(2,3)}{\so(1,3)}, 
\end{equation}
where 
\begin{equation}
\frac{\so(2,3)}{\so(1,3)}=\text{span}\left\{\EK{m}\right\},\qquad m=0,1,2,3,
\end{equation}
with 
\begin{equation}
\label{EKAdS4}
\EK{0}=\frac{1}{2}\SOM{04},\quad\EK{1}=\frac{1}{2}\SOM{14},\quad\EK{2}=\frac{1}{2}\SOM{24},\quad\EK{3}=\frac{1}{2}\SOM{34} \equiv \frac{1}{2}\ED,
\end{equation}
and 
\begin{equation}
\Str\left(\EK{m}\EK{n}\right)=\frac{1}{4}\eta_{mn},\qquad m,n=0,1,2,3.
\end{equation}
The $\so(1,3)$ generators are $\left\{\SOM{01},\SOM{02},\SOM{03},\SOM{12},\SOM{13},\SOM{23}\right\}$  and an appropriate coset representative for $\AdS{4}$ is
\begin{equation}\label{gAdS4}
g_{\AdS{4}}=\exp\left(\x{0}\Ep{0}+\x{1}\Ep{1}+\x{2}\Ep{2}\right)\exp\left(\log r\ED\right),
\end{equation}
where 
$\Ep{\mu}=\SOM{\mu3}+\SOM{\mu4}, \, \mu=0,1,2.$

The $\CP{3}$ space can be written as the coset in
\begin{equation}
\su(2)\oplus\su(4)=\su(2)\oplus\u(3)\oplus\frac{\su(2)\oplus\su(4)}{\su(2)\oplus\u(3)}.
\end{equation}
A basis for this coset is
\begin{equation}
\frac{\su(2)\oplus\su(4)}{\su(2)\oplus\u(3)}=\text{span}\left\{\EK{m}\right\},\qquad m=4,\dots,9,
\end{equation}
where
\begin{equation}
\label{EKCP3}
\begin{gathered}
\EK{4}=\ELm{11},\quad\EK{5}=\ELm{12},\quad\EK{6}=\ELm{13},\\
\EK{7}=\ELm{14},\quad\EK{8}=\EH,\quad\EK{9}=\ELm{10},
\end{gathered}
\end{equation}
and 
$\EH=\ELm{6}+\ELm{9}+\EMa{1}$,
with $\ELm{m}$ given in \eqref{supergens}. We also have
\begin{equation}
\Str\left(\EK{m}\EK{n}\right)=\frac{1}{2}\d_{mn},\qquad m,n=4,\dots,9.
\end{equation}
The generators of $\su(2)\oplus\u(3)$ are $\left\{\ET{2},\EMa{2},\EMa{3},\ELm{1},\ELm{2},\ELm{3},\ELm{4},\ELm{5},\ELm{7},\ELm{8},\ET{1},\ELm{15}\right\}$, with
\begin{equation}
\ET{1}=\ELm{6}-\ELm{9},\qquad\ET{2}=\ELm{6}+\ELm{9}+2\EMa{1}.
\end{equation}
An appropriate coset representative is then
\begin{equation}\label{gCP3}
g_{\CP{3}}=\exp\left(\vphi_1\ELm{3}+\vphi_2\EL-\psi\EMa{3}\right)\exp\left(\theta_1\ELm{2}+(\theta_2+\pi)\ELm{14}\right)\exp\left(\left(2\xi+\pi\right)\left(\ELm{10}+\EMa{2}\right)\right),
\end{equation}
where 
\begin{equation}
\label{ELs}
\EL=-\frac{1}{\sqrt{3}}\ELm{8}-\sqrt{\frac{2}{3}}\ELm{15}.
\end{equation}
Therefore, the full bosonic representative for $\AdS{4}\times\CP{3}$ is
\begin{equation}
g=g_{\AdS{4}}\times g_{\CP{3}}.
\end{equation}


In order to find out the undeformed background we will use the procedure taken in \autoref{BCG}. Now the bosonic algebra is $\so(2,3)\oplus\su(2)\oplus\su(4)$ so that the projector \eqref{P2def} becomes 
\begin{equation}
\label{P2X}
\Pb{X}=\frac{1}{4}\sum_{\mu=0}^3\frac{\Str\left(\EK{m}X\right)}{\Str\left(\EK{m}\EK{m}\right)}\EK{m}+\frac{1}{2}\sum_{m=4}^9\frac{\Str\left(\EK{m}X\right)}{\Str\left(\EK{m}\EK{m}\right)}\EK{m},
\end{equation}
and \eqref{PJK} turns into $\Pb{A}=\Em{m}\EK{m}, \, m=0,1,\dots,9$. Using the parametrization \eqref{gAdS4} and \eqref{gCP3} we find 
\begin{equation}
\begin{gathered}
\Em{0}=\frac{1}{2}r\dx{0},\quad\Em{1}=\frac{1}{2}r\dx{1},\quad\Em{2}=\frac{1}{2}r\dx{2},\quad\Em{3}=\frac{dr}{2r},\\
\Em{4}=\frac{1}{2}\sin\th{1}\cos\xi\dvf{1},\quad\Em{5}=\frac{1}{2}\cos\xi\dth{1},\quad\Em{6}=-\frac{1}{2}\sin\th{2}\sin\xi\dvf{2},\\
\Em{7}=-\frac{1}{2}\sin\xi\dth{2},\quad\Em{8}=\frac{1}{4}\left(\cos\th{1}\dvf{1}-\cos\th{2}\dvf{2}+2\dps\right)\sin 2\xi,\quad\Em{9}=\dji.
\end{gathered}
\end{equation}
Then the undeformed $\AdS{4}\times\CP{3}$ \footnote{{Henceforth, we set $R_{\text{str}}^2=R^3/k=2^{5/2}\pi\sqrt{N/k}=1$, where $R_{\text{str}}^2$ is defined in \cite{Aharony2008}.}} metric can be found from \eqref{Gmn} with $\eta=0$ so that from \eqref{J1} we have $J=A$ and then 
\begin{equation}
ds^2_{\AdS{4}}=\frac{1}{4}\left(r^2\left(-dx_0^2+dx_1^2+dx_2^2\right)+\frac{dr^2}{r^2}\right),
\end{equation}
and 
\begin{equation}
\label{CP3metric}
\begin{gathered}
ds^2_{\CP{3}}=\dji^2+\frac{1}{4}\cos^2\!\xi\left(\dth{1}^2+\sin^2\!\th{1}\dvf{1}^2\right)+\frac{1}{4}\sin^2\!\xi\left(\dth{2}^2+\sin^2\!\th{2}\dvf{2}^2\right)\\+\left(\frac{1}{2}\cos\th{1}\dvf{1}-\frac{1}{2}\cos\th{2}\dvf{2}+\dps\right)^2\sin^2\!\xi\cos^2\!\xi,
\end{gathered}
\end{equation}
where $\left(\th{1},\vf{1}\right)$ and $\left(\th{2},\vf{2}\right)$ parametrize the two spheres of $\CP{3}$, the angle $\xi$, $0\leq\xi\leq\pi/2$, determines their radii and $0\leq\psi\leq2\pi$. This background comes from the eleven-dimensional M-theory on $\AdS{4}\times\Sph{7}/\Z{k}$ corresponding to $\N=6$ $U(N)\times U(N)$ superconformal Chern-Simons theory at levels $k$ and $-k$ when we take large $N$ and $k$  \cite{Aharony2008}. Due to the $\Z{k}$ quotient the $\Sph{7}$ can be expressed as a $\Sph{1}$ fibration over $\CP{3}$. And since the radius of $\Sph{1}$ becomes small as $k$ increases the internal space reduces to $\CP{3}$ which gives the $\AdS{4}\times\CP{3}$ ten-dimensional spacetime. The $\CP{3}$ metric \eqref{CP3metric} is a particular form of the Fubini-Study metric \cite{Cvetic2001}.

\section{Yang Baxter Deformed Backgrounds \label{news}}
In this section we present some $r$-matrices satisfying the CYBE and build the corresponding $\AdS{4}\times\CP{3}$ deformed  backgrounds identifying their gravity duals. 

\subsection{Noncommutative ABJM Theory  \label{NCABJM}}
Let us first consider an Abelian $r$-matrix like
\begin{equation}
\label{rnoncomm}
r=\mu\,\Ep{1}\wedge\Ep{2},
\end{equation}
involving momenta operators on $\AdS{4}$ along $\x{1}$ and $\x{2}$ with $\mu$ the deformation parameter \footnote{The deformation parameter $\eta$ can always be absorbed in the $R$ operator in \eqref{J1} and \eqref{J2} so that it is present in $\mu$.}.
The nonzero components of $\Lmn{m}{n}$ in \eqref{Lmatrix} are 
\begin{equation}
\begin{gathered}
\Pb{\Rg{\EK{1}}}=\Lmn{1}{2}\EK{2},\quad\Pb{\Rg{\EK{2}}}=\Lmn{2}{1}\EK{1},
\end{gathered}
\end{equation}
with
\begin{equation}
\begin{gathered}
\Lmn{1}{2}=\Lmn{2}{1}=-\frac{\mu}{2}r^2,
\end{gathered}
\end{equation}
while the nonvanishing elements of $\Cmn{m}{n}$ in \eqref{Cmatrix} are
\begin{equation}
\begin{gathered}
\Cmn{0}{0}=\Cmn{3}{3}=\Cmn{4}{4}=\Cmn{5}{5}=\Cmn{6}{6}=\Cmn{7}{7}=\Cmn{8}{8}=\Cmn{9}{9}=1,\\
\Cmn{1}{1}=\Cmn{2}{2}=\M,\\
\Cmn{2}{3}=-\Cmn{3}{2}=-\frac{1}{2}\mu \M r^2,
\end{gathered}
\end{equation}
where 
\begin{equation}
\M^{-1}=1+\frac{\mu^2r^4}{4}.
\end{equation}
The deformed metric can then be obtained from \eqref{Gmn}
\begin{equation}
\label{metricNC4}
ds^2=\frac{1}{4}\left(r^2\left(-dx_0^2+\M\left(dx_1^2+dx_2^2\right)\right)+\frac{dr^2}{r^2}\right)+ds^2_{\CP{3}}.
\end{equation}
The $B$-field is obtained from \eqref{Bmn},
\begin{equation}
\label{Bnon}
B=\frac{\mu\M r^4}{4}\dx{1}\wedge\dx{2},
\end{equation}
which introduces the noncommutativity in the $(\x{1},\x{2})$-plane as $\left[\x{1},\x{2}\right]\sim\mu$ \cite{Ardalan1999,Chu1999,Maldacena1999,Seiberg1999,Alishahiha1999}. The commutative theory is recovered when $\mu=0$ since  $\M=1$ and $B=0$. 

This result agrees with the gravity dual of noncommutative ABJM obtained by TsT transformations, with $\g=2\mu$, where $\g$ is the TsT deformation parameter \cite{Imeroni2008}. Other choices for the $r$-matrix like $\Ep{0}\wedge\Ep{1}$ and $\Ep{0}\wedge\Ep{2}$ leads to noncommutativity in $(\x{0},\x{1})$ and $(\x{0},\x{2})$ directions, respectively, but they correspond to non-unitary and non-causal quantum field theories \cite{Gomis2000,Seiberg2000}. This is in agreement with the fact that
Yang-Baxter deformations of $AdS$, i.e. involving only generators of the conformal algebra, are dual to conformal twists that give rise to general noncommutative structures in the field theory side \cite{Tongeren2016,Araujo2017a,Araujo2018}.

\subsection{Dipole Deformed ABJM Theory \label{dipole}}
Let us now consider a $r$-matrix with three parameters
\begin{equation}
\label{rdipole}
r=\Ep{2}\wedge\left(\mu_1\ELm{3}+\mu_2\EL+\mu_3\EMa{3}\right),
\end{equation}
where $\EL$, defined in \eqref{ELs}, $\ELm{3}$ and $\EMa{3}$ are the Cartan generators of $\su(2)\oplus\su(4)$ and $\mu_i$, $i=1,2,3$, are the deformation parameters. In this case \eqref{rdipole} combines generators of both subspaces which will lead to a deformation of the entire $\AdS{4}\times\CP{3}$ background. Taking the same steps as in the previous case we find the nonzero components of $\Lmn{m}{n}$ 
\begin{equation}
\begin{gathered}
\Lmn{2}{4}=-\Lmn{4}{2}=-\frac{\mu_1}{4}\,r\sin\th{1}\cos\xi,\\
\Lmn{2}{6}=-\Lmn{6}{2}=\frac{\mu_2}{4}\,r\sin\th{2}\sin\xi,\\
\Lmn{2}{8}=-\Lmn{8}{2}=\frac{1}{8}\,r\left(2\mu_3-\mu_1\cos\th{1}+\mu_2\cos\th{2}\right)\sin 2\xi,
\end{gathered}
\end{equation}
as well as the nonzero elements of $\Cmn{m}{n}$ 
\begin{align}
\nonumber
\Cmn{0}{0}=&\Cmn{1}{1}=\Cmn{3}{3}=\Cmn{5}{5}=\Cmn{7}{7}=\Cmn{9}{9}=1,\\
\nonumber
\Cmn{2}{2}=&\M,\\
\nonumber
\Cmn{2}{4}=&-\Cmn{4}{2}=+\frac{1}{2}\M\mu_1 r\sin\th{1}\cos\xi,\\
\nonumber
\Cmn{2}{6}=&-\Cmn{6}{2}=-\frac{1}{2}\M\mu_2 r\sin\th{2}\sin\xi,\\
\nonumber
\Cmn{2}{8}=&-\Cmn{8}{2}=-\frac{1}{4}\M r\left(2\mu_3-\mu_1\cos\th{1}+\mu_2\cos\th{2}\right)\sin 2\xi,\\
\nonumber
\Cmn{4}{4}=&\M\left(1+\frac{r^2}{4}\mu_2^2\sin^2\!\th{2}\sin^2\!\xi+\frac{r^2}{16}\left(2\mu_3-\mu_1\cos\th{1}+\mu_2\cos\th{2}\right)^2\sin^2\!2\xi\right),\\
\nonumber
\Cmn{4}{6}=&\Cmn{6}{4}=+\frac{1}{8}\M\mu_1\mu_2 r^2\sin\th{1}\sin\th{2}\sin2\xi,\\
\nonumber
\Cmn{4}{8}=&\Cmn{8}{4}=+\frac{1}{8}\M\mu_1 r^2\sin\th{1}\left(2\mu_3-\mu_2\cos\th{1}+\mu_1\cos\th{2}\right)\cos\xi\sin2\xi,\\
\nonumber
\Cmn{6}{6}=&\M\left(1+\frac{r^2}{4}\mu_1^2\sin^2\!\th{1}\cos^2\!\xi+\frac{r^2}{16}\left(2\mu_3-\mu_1\cos\th{1}+\mu_2\cos\th{2}\right)^2\sin^2\!2\xi\right),\\
\nonumber
\Cmn{6}{8}=&\Cmn{8}{6}=-\frac{1}{8}\M\mu_2 r^2\sin\th{2}\left(2\mu_3-\mu_1\cos\th{1}+\mu_2\cos\th{2}\right)\sin\xi\sin2\xi,\\
\Cmn{8}{8}=&\M\left(1+\frac{r^2}{4}\left(\mu_1^2\sin^2\!\th{1}\cos^2\!\xi+\mu_2^2\sin^2\!\th{2}\sin^2\!\xi\right)\right),&
\end{align}
where
\begin{equation}
\begin{gathered}
\M^{-1}=1+f_1^2+f_2^2+f_3^3, \\
f_1=\frac{r}{2}\mu_1\sin\th{1}\cos\xi,\quad f_2=\frac{r}{2}\mu_2\sin\th{2}\sin\xi,\\
f_3=\frac{r}{4}\left(2\mu_3-\mu_1\cos\th{1}+\mu_2\cos\th{2}\right)\sin 2\xi.
\end{gathered}
\end{equation}
The deformed metric is now 
\begin{equation}
\label{metricdipole3}
\begin{gathered}
ds^2=\frac{1}{4}\left(r^2\left(-\dxs{0}+\dxs{1}\right)+\M r^2\dxs{2}+\frac{dr^2}{r^2}\right)\\
+\dji^2+\frac{1}{4}\cos^2\!\xi\left(\dth{1}^2+\M \left(1+f_2^2+f_3^2\right)\sin^2\!\th{1}\dvf{1}^2\right)\\
+\frac{1}{4}\sin^2\!\xi\left(\dth{2}^2+\M \left(1+f_1^2+f_3^2\right)\sin^2\!\th{2}\dvf{2}^2\right)\\+\M\left(1+f_1^2+f_2^2\right)\left(\frac{1}{2}\cos\th{1}\dvf{1}-\frac{1}{2}\cos\th{2}\dvf{2}+\dps\right)^2\sin^2\!\xi\cos^2\!\xi\\
+\M f_3\left(f_1\sin\th{1}\cos\xi\dvf{1}+f_2\sin\th{2}\sin\xi\dvf{2}\right)\left(\frac{1}{2}\cos\th{1}\dvf{1}-\frac{1}{2}\cos\th{2}\dvf{2}+\dps\right)\sin\xi\cos\xi\\
-\frac{\M}{2}f_1 f_2\sin\th{1}\sin\th{2}\sin\xi\cos\xi\dvf{1}\dvf{2},
\end{gathered}
\end{equation}
while the $B$-field is 
\begin{equation}
\begin{gathered}
B=\frac{1}{2}\M r\cos\xi\left(f_1\sin\th{1}-f_3\cos\th{1}\sin\xi\right)\dx{2}\wedge\dvf{1}\\
+\frac{1}{2}\M r\sin\xi\left(f_3\cos\th{2}\cos\xi+f_2\sin\th{2}\right)\dx{2}\wedge\dvf{2}\\
-\M r f_3\cos\xi\dx{2}\wedge\dps.
\end{gathered}
\end{equation}
It is worth mentioning that the choice of generators in \eqref{rdipole} is dictated by the place where we want put the two-tori from the TsT perspective. In the present case we have one coordinate in $\AdS{4}$ and a combination of the $U(1)$'s in $\CP{3}$. The resulting metric \eqref{metricdipole3} has  deformations along the $\x{2}$-direction in $\AdS{4}$ and along the $(\vf{1},\vf{2},\psi)$ angles in $\CP{3}$.

For $\mu_1=\mu_2=0$ and $\mu_3=\mu$ the last two lines in \eqref{metricdipole3} vanish and  we obtain 
\begin{equation}
\begin{gathered}
ds^2=\frac{1}{4}\left(r^2\left(-\dxs{0}+\dxs{1}\right)+\frac{r^2}{1+f_3^2}\dxs{2}+\frac{dr^2}{r^2}\right)\\
+\dji^2+\frac{1}{4}\cos^2\!\xi\left(\dth{1}^2+\sin^2\!\th{1}\dvf{1}^2\right)+\frac{1}{4}\sin^2\!\xi\left(\dth{2}^2+\sin^2\!\th{2}\dvf{2}^2\right)\\
+\frac{1}{1+f_3^2}\left(\frac{1}{2}\cos\th{1}\dvf{1}-\frac{1}{2}\cos\th{2}\dvf{2}+\dps\right)^2\sin^2\!\xi\cos^2\!\xi,
\end{gathered}
\end{equation}
and
\begin{equation}
\label{Bdipole}
\begin{gathered}
B=-\frac{1}{4}\left(\frac{f_3}{1+f_3^2}\right)r\,dx_2\wedge\left(\frac{1}{2}\cos\th{1}\dvf{1}-\frac{1}{2}\cos\th{2}\dvf{2}+\dps\right)\sin\xi\cos\xi,
\end{gathered}
\end{equation}
with 
\begin{equation}
f_3=\frac{\mu r}{2}\sin 2\xi.
\end{equation}
This agrees with the dipole deformed ABJM theory obtained by a TsT transformation \cite{Imeroni2008}. The $B$-field \eqref{Bdipole} does not give rise to noncommutativity in the dual field theory because it has legs in  $\AdS{4}$ as well as in $\CP{3}$. Instead, the components $B_{2i}$ for $i$ being the labels for the $\CP{3} $coordinates $\vf{1},\vf{2},\psi$, give rise to a dipole vector that introduces nonlocality in the field theory \cite{Dasgupta2000,Bergman2002}.

\subsection{Nonrelativistic ABJM Theory  \label{Sch4}}
In order to construct this deformation we must use light-cone coordinates in $\AdS{4}$. Then the coset representative is now 
\begin{equation}\label{gAdS4LC}
g_{\AdS{4}}=\exp\left(\xm\Epm+\xp\Epp+\x{1}\Ep{1}\right)\exp\left(\log r\ED\right),
\end{equation}
with
\begin{equation}
\Ep{\pm}=\frac{1}{\sqrt{2}}\left(\Ep{0}\pm\Ep{2}\right),\qquad x_{\pm}=\frac{1}{\sqrt{2}}\left(\x{0}\pm\x{2}\right),
\end{equation}
while for $\CP{3}$ we keep the same form as \eqref{gCP3}. The $\AdS{4}$ metric is then
\begin{equation}
ds^2_{\AdS{4}}=-2r^2\dxp\dxm+r^2dx_1^2+\frac{dr^2}{r^2},
\end{equation}
while the $\CP{3}$ metric is still given by \eqref{CP3metric}.

Let us now consider the $r$-matrix \eqref{rdipole} with $\Ep{2}$ replaced by $\Epm$ 
\begin{equation}
\label{rSch4}
r=\Epm\wedge\left(\mu_1\ELm{3}+\mu_2\EL+\mu_3\EMa{3}\right).
\end{equation}
The nonzero components of $\Lmn{m}{n}$ \eqref{Lmatrix} are
\begin{equation}
\begin{gathered}
\Lmn{0}{4}=\Lmn{4}{0}=\frac{\mu_1}{4\sqrt{2}}r\sin\th{1}\cos\xi,\\
\Lmn{0}{6}=\Lmn{6}{0}=-\frac{\mu_2}{4\sqrt{2}}r\sin\th{2}\sin\xi,\\
\Lmn{0}{8}=\Lmn{8}{0}=-\frac{1}{4\sqrt{2}}r\left(2\mu_3-\cos\th{1}+\mu_2\cos\th{2}\right)\sin\xi\cos\xi,\\
\Lmn{2}{4}=-\Lmn{4}{2}=\frac{\mu_1}{4\sqrt{2}}r\sin\th{1}\cos\xi,\\
\Lmn{2}{6}=-\Lmn{6}{2}=-\frac{\mu_2}{4\sqrt{2}}r\sin\th{2}\sin\xi,\\
\Lmn{2}{8}=-\Lmn{8}{2}=-\frac{1}{4\sqrt{2}}r\left(2\mu_3-\cos\th{1}+\mu_2\cos\th{2}\right)\sin\xi\cos\xi,
\end{gathered}
\end{equation}
while the nonzero elements of $\Cmn{m}{n}$ are now
\begin{align}
\nonumber
\Cmn{0}{0}&=1+\frac{r^2}{8}\left(\mu_1^2\sin^2\!\th{1}\cos^2\!\xi+\mu_2^2\sin^2\!\th{2}\sin^2\!\xi+\left(2\mu_3-\mu_1\cos\th{1}+\mu_2\cos\th{2}\right)^2\sin^2\!\xi\cos^2\!\xi\right),\\
\nonumber
\Cmn{0}{2}&=-\Cmn{2}{0}=-\frac{r^2}{8}\left(\mu_1^2\sin^2\!\th{1}\cos^2\!\xi+\mu_2^2\sin^2\!\th{2}\sin^2\!\xi+\left(2\mu_3-\mu_1\cos\th{1}+\mu_2\cos\th{2}\right)^2\sin^2\!\xi\cos^2\!\xi\right),\\
\nonumber
\Cmn{0}{4}&=\Cmn{4}{0}=\frac{r}{2\sqrt{2}}\mu_1\sin\th{1}\cos\xi,\\
\nonumber
\Cmn{0}{6}&=\Cmn{6}{0}=-\frac{r}{2\sqrt{2}}\mu_2\sin\th{2}\sin\xi,\\
\nonumber
\Cmn{0}{8}&=\Cmn{8}{0}=-\frac{r}{2\sqrt{2}}\left(2\mu_3-\mu_1\cos\th{1}+\mu_2\cos\th{2}\right)\sin\xi\cos\xi,\\
\nonumber
\Cmn{2}{2}&=1-\frac{r^2}{8}\left(\mu_1^2\sin^2\!\th{1}\cos^2\!\xi+\mu_2^2\sin^2\!\th{2}\sin^2\!\xi+\left(2\mu_3-\mu_1\cos\th{1}+\mu_2\cos\th{2}\right)^2\sin^2\!\xi\cos^2\!\xi\right),\\
\nonumber
\Cmn{2}{4}&=-\Cmn{4}{2}=\frac{r}{2\sqrt{2}}\mu_1\sin\th{1}\cos\xi,\\
\nonumber
\Cmn{2}{6}&=-\Cmn{6}{2}=-\frac{r}{2\sqrt{2}}\mu_2\sin\th{2}\sin\xi,\\
\nonumber
\Cmn{2}{8}&=-\Cmn{8}{2}=-\frac{r}{2\sqrt{2}}\left(2\mu_3-\mu_1\cos\th{1}+\mu_2\cos\th{2}\right)\sin\xi\cos\xi,\\
\Cmn{1}{1}&=\Cmn{3}{3}=\Cmn{4}{4}=\Cmn{5}{5}=\Cmn{6}{6}=\Cmn{7}{7}=\Cmn{8}{8}=\Cmn{9}{9}=1.
\end{align}
The deformed background is then
\begin{equation}
\label{metricSch4}
{ds}^2=\frac{1}{4}\left(-2r^2\dxp\dxm+r^2dx_1^2+\frac{dr^2}{r^2}-\M\,r^2\dxps\right)+ds^2_{\CP{3}},
\end{equation}
with 
\begin{equation}
\begin{gathered}
\M=f_1^2+f_2^2+f_3^2, \\
%
f_1=\frac{r}{2\sqrt{2}}\mu_1\sin\th{1}\cos\xi,\quad f_2=\frac{r}{2\sqrt{2}}\mu_2\sin\th{2}\sin\xi,\\
f_3=\frac{r}{2\sqrt{2}}\left(2\mu_3-\mu_1\cos\th{1}+\mu_2\cos\th{2}\right)\sin\xi\cos\xi.
\end{gathered}
\end{equation}
The first term in \eqref{metricSch4} is a Schr\"odinger spacetime\footnote{{The Schr\"odinger symmetry is the maximal symmetry group of the free Schr\"odinger equation. It is the nonrelativistic version of the conformal algebra \cite{Hagen1972,Niederer1972}. This symmetry is realized geometrically as Schr\"odinger spacetimes.}}.   
The $B$-field is now 
\begin{equation}
\label{BSch4}
\begin{gathered}
B=-\frac{1}{\sqrt{2}}r\cos\xi\left(f_1\sin\th{1}-f_3\cos\th{1}\sin\xi\right)\dxp\wedge\dvf{1}\\
-\frac{1}{\sqrt{2}}r\sin\xi\left(f_3\cos\th{2}\cos\xi+f_2\sin\th{2}\right)\dxp\wedge\dvf{2}\\
+\frac{1}{\sqrt{2}}r\sin 2\xi f_3\dxp\wedge\dps.
\end{gathered}
\end{equation}
The choice of generators in \eqref{rSch4} is very similar to the one in \eqref{rdipole}. Now, however, the two-tori defined by the TsT transformation takes the $\xm$ coordinate and a combination of the internal $U(1)$'s in ${\CP{3}}$ and does not introduce any noncommutativity in the dual field theory. 

The metric \eqref{metricSch4} has the form of a  Schr\"odinger spacetime with dynamical exponent two \cite{Son2008,Balasubramanian2008} \footnote{The dynamical $z$ factor is the exponent in the power of the radial direction in the $r^{2z}\dxps$ term. To have Schr\"odinger symmetry we must have $z=2$. The relativistic symmetry corresponds to $z=1$. A Schr\"odinger spacetime $\Sch{4}\times\CP{3}$ with dynamical exponent three was reported in \cite{Singh2011}.}. This type of background corresponds to gravity duals of nonrelativistic field theories in M-theory 
\cite{Ooguri2010}.
The nonrelativistic ABJM theory has its origin in  M-theory where it is described as the dual field theory of M2-branes in an orbifold space. It has been suggested that there are several nonrelativistic gravity duals with Schr\"odinger symmetry \cite{Nakayama2009}. The number of $\Sch{4}\times\CP{3}$ spaces is equal to the degeneracy of a scalar harmonic function $\Phi$ on $\CP{3}$, for which $-\Lap\Phi=\l_k\Phi$ with $\l_k=4k\left(k+3\right)$. For each $\l_k$  we have $\text{deg}(\l_k)=\frac{1}{12}(1+k)^2(2+k)^2(2k+3)$ \cite{Berger1971}. The value of $k$ is fixed by requiring that the $R_{++}$ component of the Ricci tensor, which involves $\Phi$, vanishes \cite{Hartnoll2008}. Our background has $\Phi=\M/r^2$ and $-\Lap\Phi=4\l_1\Phi$ \footnote{The factor 4 in the Laplace-Beltrami equation comes from the fact that the radius of $\CP{3}$ is twice the radius of $\AdS{4}$.}, with $\l_1=16$, which has multiplicity $\text{deg}(\l_1)=15$. This indicates that the Schr\"odinger background we obtained is one of a family of fifteen solutions. The others solutions can be obtained from the fourteen $r$-matrices of the form $r=\Epm\wedge\sum_i\nu_i\ET{i}$, where $\ET{i}$ are generators of $\su(2)\oplus\su(4)$ that are not Cartan generators and $\nu_i$ are the deformation parameters. 
The $r$-matrices generating these others $\Sch{4}\times\CP{3}$ solutions have the form $r=\mu_i\,\Epm\wedge\ELm{i}$, with $i=1,2,4,5,6,7,9,10,11,12,13,14$ and $r=\mu_a\,\Epm\wedge\EMa{a}$, with $a=1,2$. In the first case, with $i=1$ we have  
\begin{equation}
r=\mu\,\Epm\wedge\ELm{1},
\end{equation}
and we get
\begin{equation}
\begin{gathered}
{ds}^2=\frac{1}{4}\left(-2r^2\dxp\dxm+r^2dx_1^2+\frac{dr^2}{r^2}-\M\,r^2\dxps\right)+ds^2_{\CP{3}},\\
B=\frac{r\cos\xi}{\sqrt{2}}\dxp\wedge\Big(f_2\dth{1}+\left(f_1\sin\th{1}-f_3\cos\th{1}\sin\xi\right)\dvf{1}+f_3\cos\th{2}\sin\xi\dvf{2}\\
-2f_3\sin\xi\dps\Big),
\end{gathered}
\end{equation}
with 
\begin{equation}
\begin{gathered}
\M=f_1^2+f_2^2+f_3^2,\\
f_1=\frac{r}{2\sqrt{2}}\mu\cos\th{1}\cos\vf{1}\cos\xi,\quad f_2=\frac{r}{2\sqrt{2}}\mu\sin\th{1}\cos\xi,\\
f_3=\frac{r}{2\sqrt{2}}\mu\sin\th{1}\cos\vf{1}\sin\xi\cos\xi.
\end{gathered}
\end{equation}
For the second family of $r$-matrices, taking $a=2$ we have 
\begin{equation}
r=\mu\,\Epm\wedge\EMa{2},
\end{equation}
and we find 
\begin{equation}
\begin{gathered}
{ds}^2=\frac{1}{4}\left(-2r^2\dxp\dxm+r^2dx_1^2+\frac{dr^2}{r^2}-f^2\,r^2\dxps\right)+ds^2_{\CP{3}},\\
B=-\frac{fr\sin\xi\cos\xi}{\sqrt{2}}\dxp\wedge\left(\cos\th{1}\dvf{1}-\cos\th{2}\dvf{2}+2\dps\right),
\end{gathered}
\end{equation}
with 
\begin{equation}
f=\frac{r}{2\sqrt{2}}\mu \cos 2\xi\sin\psi.
\end{equation}
These deformed backgrounds are all $\Sch{4}\times\CP{3}$ and belong to the family of nonrelativistic ABJM theories mentioned above.

\section{Conclusions \label{conclusions}}

 We computed explicitly backgrounds generated by some $r$-matrices which satisfy the CYBE in deformed $\AdS{4}\times\CP{3}$. By considering an abelian Jordanian $r$-matrix 
we obtained the metric and $B$-field of the gravity dual of the non-commutative ABJM theory. 
By choosing a $r$-matrix with three Cartan generators, one in $\AdS{4}$ and a combination  of two generators in  $\CP{3}$, we obtained backgrounds for a dipole deformed ABJM theory. 
These backgrounds coincide with those obtained via TsT transformations \cite{Imeroni2008}. We also considered a $r$-matrix built in a similar way but using a light-cone component of the momenta. 
Such deformed backgrounds  
are the gravity duals of nonrelativistic ABJM theories in Schr\"odinger spacetime. These backgrounds are also expected to be obtained by an appropriate null Melvin twist \cite{Bobev2009}.  An interesting point is to 
compute the corresponding TsT null Melvin twist of the undeformed $\AdS{4}\times\CP{3}$ background since it should give the same Yang-Baxter deformed backgrounds we derived. 
All these deformations of $\AdS{4}\times\CP{3}$ lend support to the relation between TsT transformations and solutions of the CYBE known as the gravity/CYBE correspondence \cite{Matsumoto2014}. 


We have focused only on the NSNS sector of the background. To compute the RR fields we must take into account the fermionic degrees of freedom. Since the RR fields can be obtained by TsT transformations \cite{Imeroni2008} it should be straightforward to obtain them by just  including fermions in our parametrization as done for the $\AdS{5}\times\Sph{5}$ case \cite{Kyono2016}.

\appendix

\section{A Basis for the \texorpdfstring{$\so(2,3)$}{so(2,3)} Algebra \label{ApSO23}} 
The 10 generators of $SO(2,3)$ can be written as
\begin{equation}\label{gens}
m_{ij}=\frac{i}{4}\left[\Gamma_i,\Gamma_j\right],
\end{equation}
and satisfy
\begin{equation}
\left[m_{ij},m_{k\ell}\right]=i\left(\eta_{i\ell}m_{jk}+\eta_{jk}m_{i\ell}-\eta_{j\ell}m_{ik}-\eta_{ik}m_{j\ell}\right),
\end{equation}
where $i,j,k,\ell=0,1,2,3,4$. 
We choose the following representation for the $SO(2,3)$ $\Gamma_i$ matrices
\begin{equation}
\left\{\Gamma_i,\Gamma_j\right\}=2\eta_{ij}, 
\end{equation}
\begin{equation}
\Gamma_i=\left\{\begin{matrix}
i\gamma_5\gamma_a & i=a=0,1,2,3\\ 
\gamma_5=i\gamma_0\gamma_1\gamma_2\gamma_3 & i=4
\end{matrix}\right.
\end{equation}
with $\eta_{ij}=\text{diag}(-+++-)$, and $\gamma_a$ being the gamma matrices in a Dirac representation $SO(1,3)$  \cite{Fabbri2000} (see \cite{Arutyunov2008} for a different choice)
\begin{equation}
\begin{gathered}
\gamma_0=\begin{pmatrix}
 I_2 &  0  \\ 
 0   & -I_2 \\  
\end{pmatrix},\quad
\gamma_1=\begin{pmatrix}
 0 &  \sigma_3  \\ 
 -\sigma_3   & 0 \\  
\end{pmatrix},\\
\gamma_2=\begin{pmatrix}
 0 &  \sigma_1  \\ 
 -\sigma_1   & 0 \\  
\end{pmatrix},\quad
\gamma_3=\begin{pmatrix}
 0 &  \sigma_2  \\ 
 -\sigma_2 & 0 \\  
\end{pmatrix}.
\end{gathered}
\end{equation}
and 
\begin{equation}
\gamma_5=\begin{pmatrix}
 0 &  I_2  \\ 
I_2 & 0 \\  
\end{pmatrix}.
\end{equation}
From \eqref{gens}, we get
\begin{equation}
m_{ab}=\frac{1}{4}\left[\gamma_a,\gamma_b\right],\qquad  m_{a4}=\frac{i}{2}\gamma_a,\qquad a,b=0,1,2,3.
\end{equation}

In order to make explicit the conformal group let us split the indices as
\begin{eqnarray}
m_{ij}=\left\{m_{\mu\nu},m_{\mu3},m_{\mu4},m_{34}\right\},\qquad  \mu,\nu=0,1,2,
\end{eqnarray}
such that $\eta_{\mu\nu}=\text{diag}(-,+,+)$ \footnote{This is going to be the signature on the Minkowskian boundary of $\AdS{4}$.}. 
Let us also define \cite{Fabbri2000}
\begin{align}\nonumber
p_\mu&=m_{\mu4}+m_{\mu3}, \\ \nonumber
k_\mu&=m_{\mu4}-m_{\mu3}, \\
D&=m_{34}.
\end{align}
Then the conformal algebra $SO(2,3)$ is
\begin{align}\nonumber
\left[m_{\mu\nu},m_{\rho\sigma}\right]&=\eta_{\mu\sigma}m_{\nu\rho}+\eta_{\nu\rho}m_{\mu\sigma}-\eta_{\mu\rho}m_{\nu\sigma}-\eta_{\nu\sigma}m_{\mu\rho},\\
\nonumber
\left[m_{\mu\nu},D\right]&=0,\\
\nonumber
\left[D,p_\mu\right]&=-p_\mu,\\
\label{conformal4}
\left[D,k_\mu\right]&=k_\mu,\\
\nonumber
\left[k_\mu,p_\nu\right]&=2\eta_{\mu\nu}D+2m_{\mu\nu},\\
\nonumber
\left[m_{\mu\nu},p_\rho\right]&=-\eta_{\mu\rho}p_\nu+\eta_{\nu\rho}p_\mu,\\
\nonumber
\left[m_{\mu\nu},k_\rho\right]&=-\eta_{\mu\rho}k_\nu+\eta_{\nu\rho}k_\mu.
\end{align}

\section{A Basis for the \texorpdfstring{$\su(4)$}{su(4)} Algebra \label{ApSU4}}

A basis for $\su(4)$ can be constructed in terms of anti-hermitian $4\times 4$ matrices known as Gell-Mann matrices,
\begin{align}\nonumber
\lambda_1=& \begin{pmatrix}
0 &1  &0  &0 \\ 
1 &0  &0 & 0\\ 
 0& 0 &  0&0 \\ 
 0&0  &0  &0 
\end{pmatrix},   \qquad  \lambda_2=\begin{pmatrix}
0 &-i  &0  &0 \\ 
i &0  &0 & 0\\ 
 0& 0 &  0&0 \\ 
 0&0  &0  &0 
\end{pmatrix}, \qquad \lambda_3=\begin{pmatrix}
1 &0  &0  &0 \\ 
0 &-1  &0 & 0\\ 
 0& 0 &  0&0 \\ 
 0&0  &0  &0 
\end{pmatrix},
\\
\nonumber
\lambda_4=& \begin{pmatrix}
0 &0  &1  &0 \\ 
0 &0  &0 & 0\\ 
 1& 0 &  0&0 \\ 
 0&0  &0  &0 
\end{pmatrix},   \qquad  \lambda_5=\begin{pmatrix}
0 &0  &-i  &0 \\ 
0 &0  &0 & 0\\ 
 i& 0 &  0&0 \\ 
 0&0  &0  &0 
\end{pmatrix}, \qquad \lambda_6=\begin{pmatrix}
0 &0  &0  &0 \\ 
0 &0  &1 & 0\\ 
 0& 1 &  0&0 \\ 
 0&0  &0  &0 
\end{pmatrix},
\\ 
\nonumber
\lambda_7=& \begin{pmatrix}
0 &0  &0  &0 \\ 
0 &0  &-i & 0\\ 
 0& i &  0&0 \\ 
 0&0  &0  &0 
\end{pmatrix},   \qquad  \lambda_8=\frac{1}{\sqrt{3}}\begin{pmatrix}
1 &0  &0  &0 \\ 
0 &1  &0 & 0\\ 
 0& 0 &  -2&0 \\ 
 0&0  &0  &0 
\end{pmatrix}, \qquad \lambda_9=\begin{pmatrix}
0 &0  &0  &1 \\ 
0 &0  &0 & 0\\ 
 0& 0 &  0&0 \\ 
 1&0  &0  &0 
\end{pmatrix},
\\
\nonumber
\lambda_{10}=& \begin{pmatrix}
0 &0  &0  &-i \\ 
0 &0  &0 & 0\\ 
 0& 0 &  0&0 \\ 
 i&0  &0  &0 
\end{pmatrix},   \qquad  \lambda_{11}=\begin{pmatrix}
0 &0  &0  &0 \\ 
0 &0  &0 & 1\\ 
 0& 0 &  0&0 \\ 
 0&1  &0  &0 
\end{pmatrix}, \qquad \lambda_{12}=\begin{pmatrix}
0 &0  &0  &0 \\ 
0 &0  &0 & -i\\ 
 0& 0 &  0&0 \\ 
 0&i  &0  &0 
\end{pmatrix},
\\
\lambda_{13}=& \begin{pmatrix}
0 &0  &0  &0 \\ 
0 &0  &0 & 0\\ 
 0& 0 &  0&1 \\ 
 0&0  &1 &0 
\end{pmatrix},   \qquad  \lambda_{14}=\begin{pmatrix}
0 &0  &0  &0 \\ 
0 &0  &0 & 0\\ 
 0& 0 &  0&-i \\ 
 0&0  &i  &0 
\end{pmatrix}, \qquad \lambda_{15}=\frac{1}{\sqrt{6}}\begin{pmatrix}
1 &0  &0  &0 \\ 
0 &1  &0 & 0\\ 
 0& 0 &  1&0 \\ 
 0&0  &0  &-3 
\end{pmatrix}.
\end{align}
The first $8$ matrices form a basis for $\su(3)\subset \su(4)$. Furthermore, these matrices are orthogonal and satisfy
\begin{equation}
\mathrm{Tr} \left(\lambda_m \lambda_n\right)=2\delta_{mn},\qquad m=1,...,15,
\end{equation}
and commutation relations
\begin{equation}
\left[\lambda_m,\lambda_n\right]=2i f_{mn}^{p}\lambda_p.
\end{equation}
A list of non-vanishing structure constants can be found in \cite{Pfeifer2003}. In this representation the Cartan generators are given by $\lambda_3$, $\lambda_8$ and $\lambda_{15}$.

\acknowledgments
We would like to thank S. van Tongeren for valuable comments. The work of L.R. was supported by CAPES. The work of V.O. Rivelles was supported by FAPESP grants 2014/18634-9 and 2019/21281-4.

\bibliographystyle{JHEP}
\bibliography{Biblioteca}
\end{document}